\documentclass[oldversion]{aa}
\usepackage{times}
\usepackage{amssymb,amsmath}
\usepackage{mathptmx}
\usepackage{multirow}
\usepackage{natbib}
\usepackage{subfigure}
\usepackage{rotating}
\usepackage[]{fontenc}
\usepackage[latin1]{inputenc}
\usepackage[]{graphicx}

\begin{document}

\title{The black holes of radio galaxies during the ``Quasar Era'':
Masses, accretion rates, and evolutionary stage \thanks{Based on
observations collected at the Very Large Telescope of ESO. Program IDs
070.A-0545, 070.A-0229, 076.A-0684, 079.A-0617, 081.A-0468,381.A-0541,
082.A-0825, 083.A-0445, 085.A-0897}}  \author{N.~P.~H.~Nesvadba\inst{1,2},
C.~De~Breuck\inst{3}, M.~D.~Lehnert\inst{4}, P.~N.~Best\inst{5},
L.~Binette\inst{6}, D.~Proga\inst{7}}
\institute{Institut d'Astrophysique Spatiale, CNRS,
    Universit\'e Paris-Sud 11, Bat. 120-121, 91405 Orsay, France \and
email: nicole.nesvadba@ias.u-psud.fr \and European Southern
Observatory, Karl-Schwarzschild Strasse, Garching bei M\"unchen,
Germany \and GEPI, Observatoire de Paris, CNRS, Universit\'e Denis
Diderot, Meudon, France \and SUPA, Institute for Astronomy, Royal
Observatory, Blackford Hill, Edinburgh, UK \and Instituto de
Astronomía, UNAM, 04510 México, DF, Mexico \and Department of Physics
and Astronomy, University of Nevada, South Maryland Parkway,Las Vegas,
NV, USA}
\titlerunning{The black holes of HzRGs} 
\authorrunning{Nesvadba et al.}
\date{Received / Accepted }

\abstract{We present an analysis of the AGN broad-line regions of
6 powerful radio galaxies at z$\sim$2 (HzRGs), which is part of a study
of a sample of 50 HzRGs with rest-frame optical imaging spectroscopy
obtained at the VLT. In 6 galaxies we detect luminous (${\cal
L}$(H$\alpha$)$=$few$\times 10^{44}$ erg s$^{-1}$), spatially
unresolved, broad (FWHM$\ge$ 10,000 km s$^{-1}$) H$\alpha$ line
emission from the nucleus (H${\rm \alpha}$BLRs), consistent with
broad-line regions of supermassive black holes with masses of
few$\times 10^9 M_{\odot}$ and accretion luminosities of a few percent
of the Eddington luminosity. In two galaxies we also detect
H$\beta$BLRs, suggesting relatively low extinction of A$_V \sim$1 mag,
which agrees with constraints from X-ray observations. Overall, we
find H$\alpha$BLRs in $\sim$20\% of the galaxies where H$\alpha$ is
observed. By relating black hole and bulge mass, we find a possible
offset towards higher black-hole masses of at most $\sim 0.6$ dex
relative to nearby galaxies at a given host mass, although each
individual galaxy is within the scatter of the local
relationship. If not entirely from systematic effects, this
would then suggest that the masses of the host galaxies have increased by
at most a factor $\approx 4$ since z$\sim 2$ relative to the
black-hole masses, perhaps through accretion of satellite galaxies or
because of a time lag between star formation in the host galaxy and
AGN fueling. We also compare the radiative and mechanical energy
output (from jets) of our targets with predictions of recent models of
``synthesis'' or ``grand unified'' AGN feedback, which postulate that AGN
with similar radiative and mechanical energy output rates to those found in
our HzRGs may be nearing the end of their period of active growth. We
discuss evidence that they may reach this stage at the same time as
their host galaxies.}


\maketitle
\section{Introduction}
\label{sec:introduction}
One of the most remarkable discoveries in extragalactic astronomy in
the past decade is the close relationship between the masses of
supermassive black holes and several properties of their bulges, such as
stellar mass \citep{magorrian98,haring04}, stellar velocity dispersion
\citep[``M-$\sigma$ relation''][]{gebhardt00, ferrarese02, tremaine02,
ferrarese02}, and the concentration parameter \citep{graham01}.
Theoretical arguments suggest that the most fundamental of these
may be a relationship that with the binding energy of the bulge
\citep{hopkins07,younger08} or perhaps with the dark-matter halo
\citep{booth09}.

A popular way to interpret these relationships is by postulating that
supermassive black holes may regulate their own growth, as well as that
of their host galaxies, by coupling a small fraction of their energy
output to the surrounding gas. This heats and unbinds significant
fractions of the gas, by inhibiting star formation \citep[``AGN
feedback'', e.g.][]{silk98, springel05, bower06, croton06, cattaneo07,
ciotti07, sijacki07, merloni08, ciotti09, fanidakis09}. The scaling
relations of black hole mass with the bulge parameters are a major
observational constraint to gauging the ``efficiency'' of this process,
the fraction of the energy output of the AGN that is being injected
into the interstellar medium of the host galaxy \citep[typically a
few$\times$0.1\%, e.g.][]{dimatteo05, merloni08, shankar08,
koerding08}.

A large number of observational studies have been dedicated to finding
direct observational constraints for the (co-)evolution of black holes
and their host galaxies across cosmic time \citep[e.g.,][]{shields03,
walter04, mclure06, peng06, shields06, woo06, treu07, alexander08,
coppin08, shen08, wang10}. Amongst those, studies of AGN at
redshifts z$\gtrsim$2 are particularly interesting, since these
redshifts correspond to the epoch when powerful AGN were most abundant
\citep[][]{pei95} and when massive galaxies evolved most quickly
\citep[e.g.,][]{caputi06}.  However, since we know little about the
underlying physics of how AGN interact with their environments and
whether feedback occurs in AGN and host galaxies of all types at these
redshifts, it is difficult to link black-hole properties and feedback
directly. This is particularly true in light of recent galaxy
evolution models, which postulate that supermassive black holes may
interact with their environments at different stages of their
evolution in different ways, where thermal feedback may dominate at
some epochs, and mechanical feedback may at others
\citep[][]{churazov05, croton06, sijacki07, merloni08, fanidakis09, ciotti09}.

Using integral-field spectroscopy of the ionized emission-line gas in
high-redshift radio galaxies at z$\ge$2 we recently identified
energetic, kpc-scale outflows of ionized gas corresponding to
significant fractions of their interstellar medium \citep[a few $\times
10^{10}$ M$_{\odot}$;][]{nesvadba06a, nesvadba07b, nesvadba08b}. These
ionized gas masses are comparable to the cold molecular gas masses in
these galaxies \citep[e.g.,][]{papadopoulos00, debreuck03, debreuck05,
nesvadba09a}, suggesting that a significant fraction of the
interstellar medium participates in the outflows. Outflow velocities
are of order of the escape velocity of massive galaxies. These may be
the observational signatures of energetic AGN-driven winds, making the
black holes of HzRGs particularly interesting targets.

Unfortunately, nuclear broad-line regions (BLRs) in HzRGs are very
rarely observed, because the central regions are obscured in these
type II AGN. This has significantly hampered observational efforts to
characterize their black-hole properties. Most notably, at redshifts
z$\ge$2, BLRs have only been anecdotally reported for 4 HzRGs
\citep[][]{rottgering97, larkin00, nesvadba06a, humphrey07}, and their
black-hole properties have not yet been fully
discussed. \citet{mclure06} studied the redshift evolution of the
black-hole bulge mass relationship for powerful radio-loud AGN out to
z$\le$2 (with 6 radio galaxies at 1.5$\le$z$\le$2.0 and one quasar at
z=2.0), but relied on the BLRs of quasars for the black-hole
properties and on radio galaxies for the host properties, assuming
that the only difference between radio-loud quasars and radio galaxies
in a given redshift range was in their orientation. This has not yet
been proven for high-redshift AGN, where evolutionary effects may be
at least as important \citep[e.g.][]{hopkins06}.

Here we discuss the properties of supermassive black holes in 6
powerful radio galaxies at z$\sim$2-2.5 (HzRGs), in which we detect
broad H$\alpha$ line emission with FWHM$\ge$10,000 km s$^{-1}$
originating in a compact region near the nucleus. The similarity to
broad-line quasars suggests this is nuclear broad-line emission from
the AGN. The emission lines are luminous, but less so than in
powerful, optically selected quasars, making it possible to study AGN
and host properties for the same objects. We discuss black-hole masses
and accretion rates, and use the stellar mass estimates of
\citet{seymour07} to place these galaxies on the black-hole bulge mass
relationship.  We also compare radiative and mechanical energy output
rates with predictions from recent AGN feedback models
\citep[][]{churazov05, sijacki07, merloni08, fanidakis09}, which
suggest that black holes with similar properties as those in our
targets may be in a critical transition phase from radiatively
efficient to radiatively inefficient accretion.

Throughout the paper we adopt a flat H$_0$= 70 km
s$^{-1}$ Mpc$^{-1}$ concordance cosmology with $\Omega_M=0.3$ and
$\Omega_{\Lambda}=$0.7.

\section{Observations and sample selection}
\label{sec:observations}
All data were taken with the near-infrared integral-field spectrograph
SINFONI \citep{tecza00,bonnet04} at the Very Large Telescope of ESO in
the H and K bands. SINFONI is an image slicer giving a field of view
of 8\arcsec$\times$8\arcsec\ with a pixel scale of 250 mas in the
seeing-limited mode. These observations are part of an ongoing survey
of HzRGs at z$\gtrsim$2.  Our sample currently consists of 50 powerful
radio galaxies at z$\gtrsim$2, which were selected according to radio
power and radio size. All targets have steep-spectrum radio
sources. They include a significant fraction of the known radio
galaxies with confirmed redshifts at z$\ge$2 observable from the VLT
and at suitable redshifts for near-infrared observations. All are
type-2 AGN, classified by narrow rest-frame UV emission lines and a
high HeII/CIV line ratio.  The results of the full survey will be
discussed in a series of upcoming papers. Here we focus on the nuclear
properties of the 6 galaxies that have broad H$\alpha$ line emission
with properties consistent with originating in AGN broad-line
regions. We defer a full analysis of the spatially-extended line
emission to a subsequent paper, but emphasize that the present
galaxies are consistent with having outflows very similar to those
described by \citet{nesvadba06a,nesvadba07b,nesvadba08b}.

Most data were obtained with the H$+$K grating giving a spectral
resolving power R$=$1500 and a spectral coverage between 1.45$\mu$m
and 2.4$\mu$m. Total on-source exposure times were between 45 min and
6 hrs per source. MRC1138-262 was observed in the H and the K band
individually, with R$\sim 3000$ and 4000, respectively, and with
on-source exposure times of 75 and 140 min, respectively \citep[The
data of MRC1138-262 have previously been described by ][but without a
dedicated analysis of its nuclear properties.]{nesvadba06a} We reduced
all data sets with our custom data reduction software, which is
optimized to study high-redshift galaxies. For details of the data
reduction procedure, see, e.g., \citet{nesvadba06a,nesvadba08}. Data
were calibrated using nearby bright stars observed at similar air
masses to the targets at the beginning or end of each observation. We
used the same stars to monitor the size of the seeing disk with
SINFONI and the same setup as for the data. Typically, the seeing was
between 0.5\arcsec\ and 1.5\arcsec, corresponding to a spatial
resolution of 4$-$12 kpc at a redshift z$\sim$2.5.

Three of our sources have already been known to have broad nuclear
H$\alpha$ line emission, but their black-hole properties have not yet
been discussed in the literature. \citet{larkin00} discovered a broad
nuclear H$\alpha$ line in MRC2025-218 at z$=$2.6. \citet{nesvadba06a}
identified a broad H$\alpha$ emission line in MRC1138-262 at
z=2.2. \citet{humphrey07} report on the discovery of broad nuclear
H$\alpha$ emission in MRC1558$-$003 at z=2.5. For MRC1558$-$003 and
MRC2025$-$218 our measurements are consistent with the previous
results. For TXS1113-178, \citet{rottgering97} find a broad CIV
component consistent with the FWHM of our H$\alpha$ line
measurement. \citet{eales96} report the detection of a ``broad''
H$\alpha$ line in MRC0156-252; however, their spectrum has low
signal-to-noise and they give an FWHM estimate of about 4000 km
s$^{-1}$.  This is consistent with the integrated extended,
spatially-resolved line emission in our data, and significantly
narrower than the FWHM$=$12500 km s$^{-1}$ we find for the broad
line. In the following we use our new measurements, adopting the
results published in \citet{nesvadba06a} for MRC1138-252. A few broad
lines were previously detected in radio galaxies at lower redshifts,
z$\sim1-1.5$ \citep{economou95,willott00}.

\section{Results}
\label{sec:results}
\begin{figure}
\centering
\includegraphics[width=0.6\textwidth]{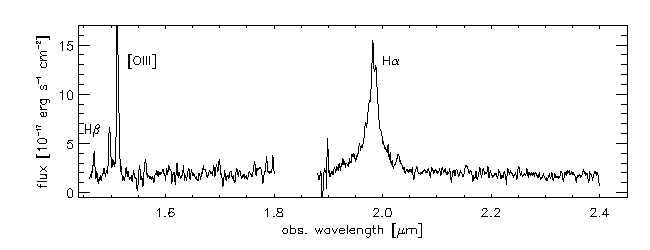}\\
\includegraphics[width=0.6\textwidth]{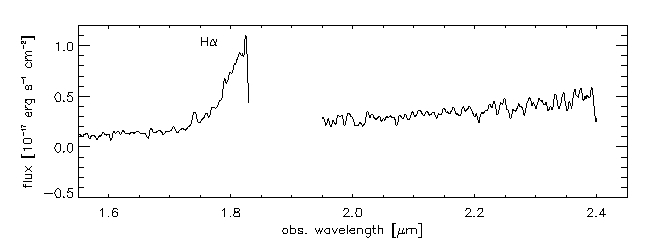}\\
\includegraphics[width=0.6\textwidth]{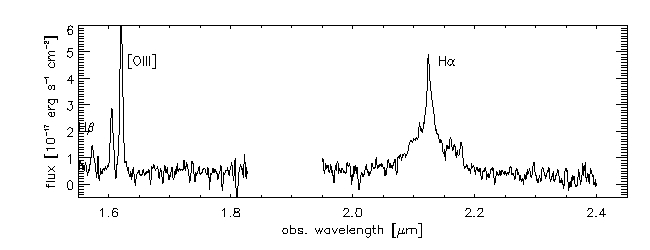}\\
\includegraphics[width=0.6\textwidth]{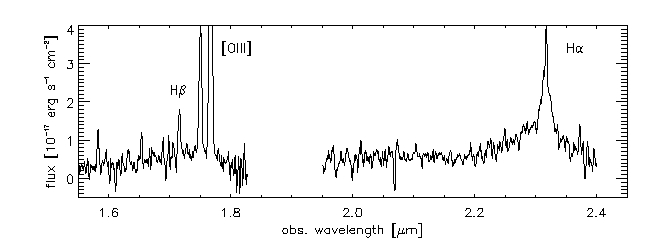}\\
\includegraphics[width=0.6\textwidth]{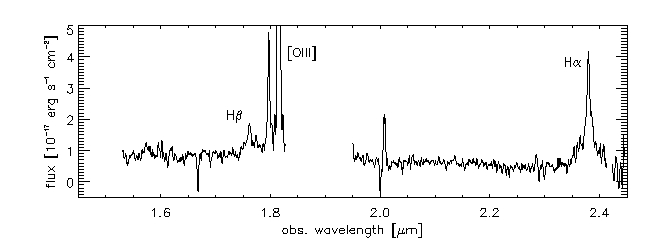}\\
\caption{Spectra of 5 of the HzRGs with BLRs seen in H$\alpha$ and SINFONI
  observations \citep[for the spectrum of MRC1138-262 see][]{nesvadba06a}. {\it
    Top to bottom:} MRC0156-252, MRC1017-220, TXS1113-178, MRC1558-003, and
  MRC2025-218. The BLR of MRC0156-252 and MRC1558-003 were detected through
  longslit spectroscopy by \citet{larkin00} and \citet{humphrey07},
  respectively. All spectra are smoothed by 5 pixels (corresponding to 25\AA\
  or $\sim$340 km s$^{-1}$) to emphasize the BLR. This is only for the sake of
  visualization, the analysis was performed on the unsmoothed spectra.}
\label{fig:spectra}
\end{figure}

Figure~\ref{fig:spectra} shows the integrated spectra of our targets
centered on the position of the brightest continuum emission. Spectra
were extracted from 2\arcsec$\times$2\arcsec\ box apertures
(corresponding to about twice the FWHM of the seeing disk). Broad
emission-line components are apparent in all six galaxies, and are
superimposed by luminous, narrow components of H$\alpha$ and
[NII]$\lambda\lambda$6548,6583. MRC1017-220 is at a redshift where the
red wing of the broad line falls into a spectral region with very low
atmospheric transmission between the H and the K band, so that we only
detected the blue wing of the broad line and the line center, but missed
most of the red wing.

Owing to the relatively broad widths of the narrow lines of typically
FWHM$\sim$800 km s$^{-1}$, [NII] and the narrow H$\alpha$ component are blended
in all galaxies. The kinematics of these lines is consistent with
those in the extended emission-line regions, suggesting they are part
of the same outflows as seen in projection along the same line of sight
as the AGN. We will discuss the extended line emission in a
forthcoming paper.

\subsection{Broad H$\alpha$ line emission} 
\label{ssec:broadha}
To derive emission-line properties, we removed the continuum emission
and fitted all broad and narrow line components with a set of weighted
Gaussian profiles. To distinguish between broad and narrow
components, we attributed zero weights to wavelengths where the narrow
components (or night-sky lines) dominate\footnote{For MRC1017-220 we
only fitted the blue wing, so that we are not able to account for line
asymmetries. By analogy with the other targets, and given the large
astrophysical uncertainties, we do not think that this has a major
impact on the results.}. We subtracted the broad components, and fitted
the residuals with lines whose redshifts and line widths were tied to
the values observed in [OIII]$\lambda$5007. \citet{nesvadba08b} find
that, in the extended emission-line regions of HzRGs, all bright,
narrow emission lines typically have very uniform kinematic
properties, at least to the precision of the observations, and we find
the same in the present analysis.

\begin{figure}
\centering
\includegraphics[width=0.7\textwidth]{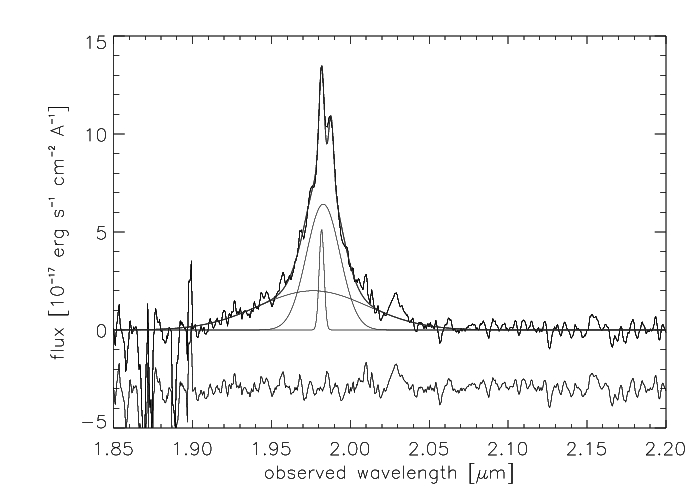}\\
\caption{Close-up of the H$\alpha$ spectral region of
MRC0156-003. We show the spectrum with the best fit superimposed
(thick black line). Thin black lines show each individual fit
component. The lower panel shows the fit residual. The line at
$\lambda$=2.03$\mu$m is the [SII]$\lambda\lambda$6716,6731 doublet,
where the two components are blended. At wavelengths shortward of
1.9$\mu$m the atmospheric transmission drops to $<$10\%.}
\label{fig:MRC0156_BLRfit}
\end{figure}

Broad H$\alpha$ lines typically have fluxes 
$F_{H\alpha}\sim 10^{-14}-10^{-15}$ erg s$^{-1}$ cm$^{-2}$ and 
FWHMs$\sim$10000-14000 km s$^{-1}$. Values for individual targets are
given in Table~\ref{tab:observations}. We did not correct for Galactic
foreground extinction, which is negligible at the observed
near-infrared wavelengths. One broad and one narrow component were
sufficient for all galaxies except MRC0156-252, for which we needed an
additional H$\alpha$ component of intermediate width, FWHM=7480 km
s$^{-1}$ and an integrated line flux of $1.4\times 10^{-14}$ erg
s$^{-1}$ cm$^{-2}$ (Figure~\ref{fig:MRC0156_BLRfit}).

All data sets are deep enough that we would have detected H$\alpha$
lines with similar luminosity and width, which suggests that HzRGs
with H$\alpha$BLRs may have different AGN properties or may be more
highly extincted than other HzRGs. To quantify this we 
measured the r.m.s. at the spectral position of H$\alpha$ in each of
the cubes where H$\alpha$ falls into the band, but was not
detected. We find a typical r.m.s. of a few$\times 10^{-19}$ erg
s$^{-1}$ cm$^{-2}$, and up to $\sim 1\times 10^{-18}$ erg s$^{-1}$
cm$^{-2}$ for the shallowest data sets. We used Monte Carlo
simulations to estimate 3$\sigma$ limits on the integrated flux of
these lines, assuming two line widths which are representative of the
measured widths. For 1000 iterations,  we can exclude line fluxes
brighter than $8.7\times 10^{-17}$ erg s$^{-1}$ cm$^{-2}$ and
$2.6\times 10^{-16}$ erg s$^{-1}$ cm$^{-2}$ at 3$\sigma$ significance
for line widths of FWHM$=$9500 km s$^{-1}$ and FWHM$=$12500 km
s$^{-1}$, respectively. This is about an order of magnitude fainter
than the fluxes of the detected broad H$\alpha$ lines.

\subsection{Broad H$\beta$ line emission and extinction}
\label{ssec:hb}
\begin{figure}
\centering
\includegraphics[width=0.45\textwidth]{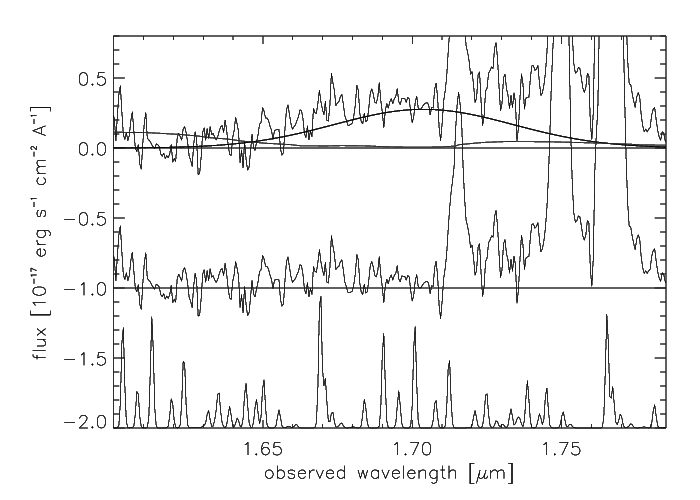}
\includegraphics[width=0.45\textwidth]{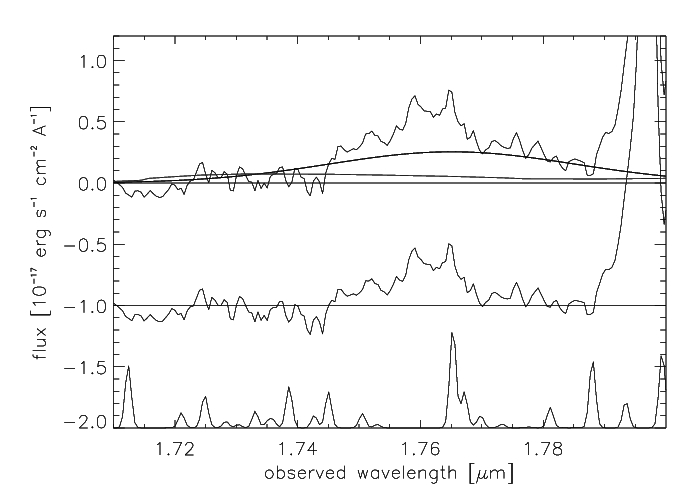}\\
\caption{We detected broad H$\beta$ line emission in two of our galaxies,
  MRC1558-003 and MRC2025-218 (upper panels). Thick black lines show a Gaussian
  line profile with the same redshift and line width, which we measured for
  H$\alpha$, scaled to match the H$\beta$ flux. Thick gray lines show the FeII
  template from \citet{iwamuro02}, to illustrate that the contribution of FeII
  is negligible in both cases. We subtracted the continuum in both
  spectra. The middle and lower panels show the fit
  residual spectrum after subtracting H$\beta$ and the position of night sky
  lines, respectively.}
\label{fig:hbspectra}
\end{figure}

In the two galaxies with the deepest observations we also detected a
broad H$\beta$ component in the H-band. Figure~\ref{fig:hbspectra}
shows the broad H$\beta$ line emission in MRC0156-252 and MRC1558-003.
TXS1113-178 and MRC1138-262 have more shallow data which are not
sufficient for probing H$\beta$ line emission in a similar way. In
MRC0156-252 and MRC1017-220, H$\beta$ falls outside the near-infrared
atmospheric windows.

Although the broad H$\beta$ lines are not intrinsically faint, they
are difficult to detect due to their large widths compared to the
resolution, imperfections in the night-sky line subtraction,
superimposed narrow-line components, including
[OIII]$\lambda\lambda$4959,5007, and the shape of the continuum. We
therefore did not fit the redshift and widths of these lines
individually, but used the values found for H$\alpha$ to identify the
lines and to estimate their fluxes. In Figure~\ref{fig:hbspectra} we
show the expected H$\beta$ profile for both galaxies. By scaling this
spectrum we derive H$\alpha$/H$\beta$ flux ratios of
F(H$\alpha$)/F(H$\beta$)$\sim$4-5 for both galaxies. By comparing our
spectra with the FeII quasar template of \citet{iwamuro02} we can rule
out strong FeII contamination at the wavelength of H$\beta$.

For an intrinsic line ratio of H$\alpha$/H$\beta=$3.1 and a galactic
extinction law, the observed H$\alpha$/H$\beta$ ratios correspond
to an extinction of A$_{\rm V}\sim$1 mag.  \citet{dong08} report very
similar H$\alpha$/H$\beta$ ratios in radio-loud AGN at low redshift,
H$\alpha$/H$\beta=$3.4.

Equally low extinctions are also suggested by comparing the observed
H$\alpha$ and X-ray fluxes. If most of the H$\alpha$ line emission is
due to photoionization of gas in the central few parsecs of the AGN,
then simple energy conservation arguments imply that the H$\alpha$
line emission cannot be more luminous than the X-ray emission. In
fact, observations of nearby AGN suggest X-ray/H$\alpha$ ratios of
3-20 \citep{ward88,imanishi04}. X-ray luminosities are available in
the literature for two of our sources. \citet{overzier05} measure
${\cal L}_X=2\times 10^{45}$ erg s$^{-1}$ at 2-6 keV for MRC0156-003,
and \citet{carilli02} find ${\cal L}_X=4\times 10^{45}$ erg s$^{-1}$
at 2-10 keV for MRC1138-262. These X-ray and H$\alpha$ luminosities
given in Table~\ref{tab:results} imply a ratio of X-ray to
H$\alpha$ emission of $\sim 6$ for both sources. This is within the
range of values suggested by \citet{ward88} and \citet{imanishi04},
and is fully consistent with relatively low extinction along the line
of sight into the AGN. \citet{alexander08} give similar arguments for
submillimeter-selected quasars and point out that this also suggests
that scattering is unlikely to play a large role for the observed
H$\alpha$ line fluxes. The same applies to our HzRGs.

The same comparison for two HzRGs with similar X-ray and near-infrared
data \citep{overzier05,nesvadba08b} suggests that the H$\alpha$BLRs of
HzRGs may generally suffer much higher extinction. \citet{overzier05}
measure ${\cal L}_X=1\times 10^{44}$ erg s$^{-1}$ for MRC0406-244 at
z=2.4 and ${\cal L}_X=6\times 10^{44}$ erg s$^{-1}$ for TXS0828+193 at
z=2.6. \citep{nesvadba08b} did not reveal any H$\alpha$BLRs but allow
us to place upper limits on their observed H$\alpha$ luminosity of
$2\times 10^{42}$ erg s$^{-1}$. This corresponds to much smaller
fractions of their X-ray luminosity, 1/20 for MRC0406-244 and 1/100
for TXS0828+193. It may indicate that extinction depends strongly on
the individual sightline into the AGN. This is plausible if the
spatial distribution of the obscuring material is inhomogeneous and
clumpy.

\section{The supermassive black holes of HzRGs}
\label{sec:bhsinhzrgs}
\subsection{The frequency of H$\alpha$BLRs in HzRGs}
\label{ssec:frequency}
Previous discoveries of H$\alpha$BLRs in HzRGs have only been anecdotal
\citep[][see also \citealt{economou95,willott00} at somewhat lower
redshifts, z$\le$1.5]{larkin00,nesvadba06a,humphrey07}. Although we
did not select a complete sample in the usual statistical sense, our
targets include a significant fraction of the known HzRGs at these
redshifts, and may therefore allow a rough assessment of how frequent
H$\alpha$BLRs may be among the most powerful radio galaxies known at
z$\ge$2. Having 27 targets at redshifts where H$\alpha$BLRs fall into
the atmospheric windows and adding 4 HzRGs at similar redshifts of
\citet{humphrey07} which are not in our sample, we find that
H$\alpha$BLRs were detected in 6/31 possible cases, corresponding to
$\sim$ 20\% of our sample.

\subsection{The nature of H$\alpha$BLRs}
\label{ssec:nature}
HzRGs typically have luminous, extended line emission with complex
kinematics. For example, \citet{nesvadba06a, nesvadba07b, nesvadba08b}
find line widths of up to FWHM$\sim$4000 km s$^{-1}$ in the integrated
spectra of HzRGs, which are due to spatially extended gas with large
velocity gradients and line widths, and could mimic a BLR (see the
example of MRC0156-262,\S\ref{sec:observations}). In our targets the
H$\alpha$ emission-line morphology shows a luminous, nuclear, and
compact H$\alpha$ emission-line component in addition to the more
diffuse extended (and narrower) line emission. These broad
components have no equivalent in [OIII]$\lambda$5007, unlike the
emission from the diffuse gas, where the spectral profiles of
H$\alpha$ and [OIII]$\lambda$5007 are typically very similar
\citep{nesvadba07b,nesvadba08b}. The absence of broad components of
forbidden lines is among the prime characteristics of BLRs where the
electron densities are high enough that forbidden levels are
collisionally de-excited.

We take this as evidence that the H$\alpha$BLRs of our targets indeed
trace gas in the vicinity of the supermassive black hole,  and
adopt the common interpretation that this gas is approximately in
virial equilibrium and is an adequate tracer of the black hole
mass. Accretion rates are relatively low in our sources
(\S\ref{ssec:accretion}) well below the values for which we may expect
strong disk winds \citep{kurosawa09,proga04}.

\subsection{Virial mass estimates}
\label{ssec:masses}
A number of recipes have been proposed to estimate
black-hole masses from the line widths and luminosities
\citep[e.g.,][]{mclure02, vestergaard02, onken04, peterson04,
peterson06}, among which we choose the approach of \citet{greene05}
which relies on the properties of the H$\alpha$ line only. This is the
obvious method of choice, because H$\alpha$ is the only line that is
well detected in our targets. 

\citet{greene05} give a simple scaling for black-hole mass, $M_{BH}$,
as a function of the FWHM and luminosity of the H$\alpha$ line,
$M_{BH} = 2\times 10^6\ {\cal L}_{42}^{0.55} \times FWHM^{2.06}_{3}$,
where FWHM$_{\rm 3}$ and H$\alpha$ luminosity ${\cal L_{\rm 42}}$ are
given in units of $10^3$ km s$^{-1}$ and $10^{42}$ erg s$^{-1}$.  With
this scaling and the luminosities and FWHMs given in
Table~\ref{tab:observations}, we find a relatively narrow range of
black-hole masses of a few $\times 10^9$ M$_{\odot}$ and up to
$2\times10^{10}$ M$_{\odot}$. Results for individual galaxies are
given in Table~\ref{tab:results}.

Applying a method calibrated for quasars to radio galaxies can
potentially introduce important systematic uncertainties, in
particular owing to inclination and extinction. Section~\ref{ssec:hb}
suggests that extinction does not likely play a major
role. \citet{greene05} find an average observed ratio of
H$\alpha$/H$\beta \approx$3.5, whereas we find H$\alpha$/H$\beta
\approx$4. Since black-hole mass depends only weakly on the H$\alpha$
luminosity, this difference is negligible.

Orientation is perhaps a bigger concern, if our radio galaxies are on
average more strongly inclined relative to the line of sight than
quasars \citep[ e.g.,][]{antonucci93}. \citet{greene05} do not find any
significant difference between radio-quiet and radio-loud quasars, but
it is likely that their radio-loud sample consisted mostly of
flat-spectrum quasars (although they do not comment on the radio
spectral index).

\citet{jarvis02} compare quasars with flat and steep-spectrum radio
sources and find on average a factor 1.3 broader lines in
steep-spectrum quasars similar to our HzRGs, which would correspond to
a bias towards higher masses by a factor 1.8.
Of course we cannot exclude that HzRGs may on
average be even more strongly inclined than steep-spectrum radio-loud
quasars. If, for example, the BLR in such a 
quasar had an inclination of $\sim$ 45$^{\circ}$ relative to the line
of sight, and that of a radio galaxy an inclination of $\sim$
80$^{\circ}$, then we would expect to overestimate the black hole mass
by another factor $\sim 2$, which is not very different from other
astrophysical uncertainties \citep[see also, e.g.,][]{collin06}.

Orientations near the line of sight are also disfavored by relatively
uniform and moderate fractions of the luminosity of the radio core
relative to the lobes of 0.7\% to 10\% \citep{carilli97,pentericci00},
which suggest orientations between 45$^\circ$ and 60$^\circ$
\citep[e.g..][]{carilli97}. This also suggests that orientation probably
does not introduce a major bias.  Moreover, it is not clear whether
the unified model is appropriate for high-redshift AGN, since
evolutionary effects may at least play as large a role at high
redshift \citep[e.g.][see also
\S\ref{sec:discussion.accretion}.]{hopkins06}. When necessary 
in the following we state both mass estimates, the one with and the one
without inclination correction.

\subsection{Eddington luminosities, radiative and mechanical luminosities}
\label{ssec:accretion}
The accretion luminosity of a black hole is the second fundamental
parameter we estimate from our data, by relating the bolometric
luminosity with the maximally possible accretion luminosity for
a spherically symmetric flow (Eddington luminosity). The Eddington
luminosity, ${\cal L}_{Edd}$, of a SMBH scales with black-hole
mass as ${\cal L}_{\rm Edd} = 1.3\times 10^{38}\ M / M_{\odot}$ erg
s$^{-1}$ \citep{rees84}.  In Table~\ref{tab:results} we list the
Eddington luminosities corresponding to the black-hole masses
estimated in \S\ref{ssec:masses}.

The bolometric luminosity can be estimated from our
data. \citet{greene05} calibrated the broad H$\alpha$ luminosity as a
function of the continuum luminosity at 5100 \AA, ${\cal L}_{H\alpha}=
5.25\times 10^{42}\ {\cal L}_{5100,44}^{1.157}$,  and we adopt
${\cal L}_{\rm bol} = 9\ \lambda\ {\cal L}_{\rm 5100}$ to translate
the continuum luminosity at 5100 \AA\ to the bolometric luminosity
\citep{kaspi00}.
With the black-hole properties listed in Table \ref{tab:results}, we
find accretion luminosities of 0.02$-$0.07 ${\cal L}_{Edd}$
(0.04$-$0.14 ${\cal L}_{Edd}$ for inclination-corrected black-hole
masses, \S\ref{ssec:masses}). This result is not dominated by
extinction. H$\alpha$/H$\beta$ line ratios are not much higher than
those of quasars \citep{greene05}, and Eddington luminosity and
bolometric luminosity only weakly depend on ${\cal L}_{H\alpha}$ with
powers of 0.55 and 0.86, respectively. Therefore extinction is 
unlikely to lower the Eddington ratios by more than $\sim 10$\%.

\subsection{Kinetic luminosity of the radio source}
\label{ssec:radio}
In \S\ref{sec:discussion.accretion} we rely on the radiative and
mechanical energy output rates of the supermassive black holes. This
requires constraining the kinetic jet luminosity, which is inherently
difficult, because the observed radio luminosity is not a good tracer
of the energy content of the relativistic jet plasma. Consequently,
jet kinetic powers are only known at an order-of-magnitude level or
worse \citep[e.g.,][]{bicknell95}. We adopt two different methods
to estimate the kinetic power in our sources. Both methods rely on
low-frequency fluxes, which we estimate from extrapolating along the
power law of the radio synchrotron emission based on multi-frequency
observations listed in the NED extragalactic data base.

First, we use the calibration of \citet{birzan08}, who estimated the
energy necessary to inflate cavities in the diffuse hot gas of massive
galaxy clusters. This is a very direct empirical estimate of the jet
kinetic power, but was derived for radio sources that are 2-3 orders
of magnitude less powerful than those in our targets. Second, we rely
on the results of \citet{willott99}, who propose a scaling of the
kinetic power as a function of the monochromatic radio flux at
151~MHz, ${\cal L}_{kin}=3\times 10^{45}\ f_W^{3/2}\ {\cal
L}_{151,28}^{6/7}$ erg s$^{-1}$. Here ${\cal L}_{151,28}$ indicates the
measured flux density at 151~MHz in units of $10^{28}$ W Hz$^{-1}$
sr$^{-1}$, and $f_W$ is a correction factor which includes the most
salient astrophysical uncertainties. It is likely in the range 1$-$20,
and we adopt a value of 10, following the arguments given in
\citet{cattaneo09}.

Typically we find kinetic luminosities of $5-10\times10^{45}$ erg
s$^{-1}$ with the calibration of \citet{birzan08}, and $\sim
5-10\times10^{46}$ erg s$^{-1}$, corresponding to of order $0.01-0.1
{\cal L}_{Edd}$ with that of \citet{willott99}. Results for individual
galaxies are listed in Table~\ref{tab:results}.

\section{Black-hole bulge mass relationship} 
\label{sec:bhbulgemass}

The scaling relationships of black holes and bulges are important
tracers of the (possibly self-)regulated growth of black holes and
galaxies across cosmic time. Accordingly, much observational effort
has been invested in tracing the redshift evolution of these
relationships \citep[e.g.,][]{shields03, shields06, woo06, peng06,
  treu07, shen08}, although sometimes with contradictory
results. Overall these observations advocate a relatively mild redshift
evolution of a few tenths of dex out to moderate redshifts (z$<<$1).

We now compare our HzRGs with the local relationship, relying on
our black-hole mass estimates (\S\ref{ssec:masses}) and the
photometric mass estimates of \citet{seymour07}.
\citeauthor{seymour07} estimated masses or upper limits for 70 HzRGs with
Spitzer rest-frame near-infrared observations, constraining the
relative contribution of the stellar component and dust at different
temperatures. Photometric estimates of stellar masses are relatively
robust for a range of star-formation histories, but poorly constrained
ages and initial mass functions cause systematic uncertainties of
factors of up to about 2-3 for plausible ages between few $\times
10^8$ yrs and about $10^9$ yrs. 5 of our galaxies are included in the
sample of \citet{seymour07}.
For TXS1113-178, which is not part of
this sample, we use the average mass of the \citeauthor{seymour07}
sample, $10^{11.5}$ M$_{\odot}$. We prefer to use this average mass
rather than the K-bands magnitude derived from our SINFONI cube,
because the latter may be significantly contaminated by the AGN,
whereas the \citeauthor{seymour07} observations are near the
rest-frame H-band where the stellar contribution to the overall
spectral energy distribution is maximal. Moreover, the HzRGs of
\citet{seymour07} span a narrow mass range overall, not much greater
than the uncertainties in the mass estimates of individual
galaxies. In the following we also assume that these are
spheroid-dominated systems, consistent with the morphological studies
of \citet{vanbreugel98} and \citet{pentericci01}, and plausible for high stellar
masses of few $10^{11}$ M$_{\odot}$.

\begin{figure}
\centering
\includegraphics[width=0.8\textwidth]{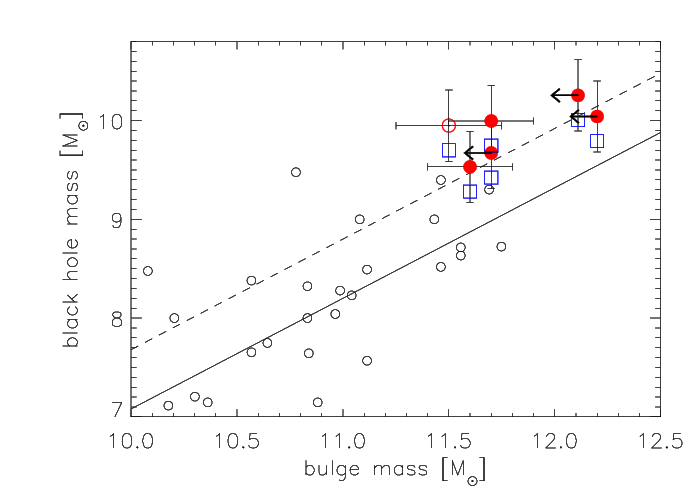}\\
\caption{Black-hole masses as estimated in \S\ref{ssec:masses} as a
function of stellar mass of the host galaxy, taken from
\citet{seymour07}. Red dots mark the HzRG in which we detected
H$\alpha$BLRs, not correcting for inclination, and blue empty squares
show the same galaxies with black-hole masses estimated assuming the
inclination correction of \citet{jarvis02}. The empty circle marks
TXS1113-178 for which we used the average HzRG mass of
\citet{seymour07}. Arrows mark galaxies for which
\citeauthor{seymour07} only provide upper limits (see text for
details). Empty circles show the nearby galaxies of \citet{haring04},
and the black dashed line shows the best linear fit to their
sample. }
\label{fig:msig}
\end{figure}

In Figure~\ref{fig:msig} we show the black-hole bulge mass
relationship of our HzRGs and compare it with the relationship for nearby
galaxies derived by \citet{haring04}. Arrows mark galaxies for
which \citet{seymour07} give upper limits on the stellar mass. A new,
ongoing analysis of the \citeauthor{seymour07} sample suggests that
these upper limits are at most 20\%-50\% larger than the actual
masses. This is negligible for our purposes. The sole exception may be
MRC1138-262 where the distinction between AGN and stellar continuum is
more difficult due to the particularly powerful AGN.  This analysis is
based on IRAC and MIPS 24$\mu$m and 60$\mu$m photometry for all
targets and includes a more systematic analysis of the dust emission
 (De Breuck et al. in prep.)

We focus in the following on the ensemble properties of our sample,
but each individual galaxy falls within the scatter of the local
relationship (Figure~\ref{fig:msig}). This illustrates that HzRGs have
at most about 4$\times$ larger black-hole masses for a given bulge
mass than expected from the local relationship (a factor 2 if we use
inclination-corrected black-hole mass estimates, see
\S\ref{ssec:masses}). This is broadly consistent with the redshift
evolution of radio-loud AGN found by \citet{mclure06}, if we
extrapolate to the redshifts of our sources. \citeauthor{mclure06}
used the observed K-band magnitudes of radio galaxies to estimate
bulge masses and steep-spectrum quasars out to z$=$1.5$-$2.0, and
predict an offset from the local relationship by 0.6 dex for z$\ge$2.

The offset we find is significantly less than those reported for
other populations of massive, strongly star-forming galaxies at
z$\ge$2, \citep[e.g.][]{walter04, alexander08, coppin08, wang10}. The
reason could be that we used stellar masses instead of dynamical
masses based on the width of CO line emission like the above analyses,
or it could indicate evolutionary effects. Redshifts z$\ge$1-2 are the
most rapid phase in the evolution of massive galaxies, which makes
observations at these redshifts particularly interesting, in spite of
the observational challenges.

The offset of HzRGs relative to the local sequence is formally
consistent with observations of the redshift evolution of the
M-$\sigma$ relationship out to moderate (z$\sim$0.5) redshifts.
Using, e.g., the relationship of \citet{woo08}, we would expect an
offset by a factor $\sim 2$. However, it is not clear whether this is
indeed an evolutionary effect or caused by selection effects. For
example, our HzRGs, which are somewhat more massive than the local
sample of \citeauthor{haring04}, would fall on the relationship by
\citet{haring04}, if the latter was somewhat steepened (within their
scatter).  Alternatively, e.g., \citet{wyithe06} propose that black
hole and bulge properties could be related through a log-quadratic
rather than a log-linear relationship, in which case our targets would
show no sign of evolution.

Observational biases may also contribute. For
example, following \citet{lauer07} we may expect that the intrinsic
scatter in the relationship combined with the the steep slope of the
high-mass end of the galaxy mass function may bias the relationship
for massive galaxies towards higher black hole masses. We argue in
\S\ref{sec:implications} that, whether this offset is intrinsic or
not, mild offsets of factors 2-4 are not inconsistent with the basic
picture where HzRGs are massive galaxies that have completed most of
their growth by z$\sim$2.

\section{Low accretion rates of SMBHs in HzRGs}
\label{sec:discussion.accretion}

In \S\ref{ssec:accretion} we used our black-hole mass estimates and
broad-line luminosities to estimate accretion luminosities for our
targets in the range of a few and up to about 10\% Eddington. When
reaching similar accretion luminosities of a few percent Eddington,
Galactic ``miniquasars'' undergo rapid changes in their accretion
properties, with dramatic effects for their X-ray and radio emission
\citep[e.g.,][]{nowak95, maccarone03, koerding06b, koerding06a,
fender04}. The analogy with these stellar-mass black holes has very
recently inspired several authors to propose a fairly complex
scenario of ``synthesis'' \citep{merloni08} or ``unified''
\citep{fanidakis09} AGN feedback, where an initial, strongly radiative
phase with high accretion rates is followed by a later phase with low
accretion rates, during which the radio source inhibits gas cooling
and accretion through the injection of kinetic energy into the ambient
medium \citep[see also, e.g.,][]{sijacki07}. \citet{churazov05}
proposes that this may be accompanied by episodes of highly unstable
accretion, rapid ``radio flaring'', and powerful radio activity. 
\citet{merloni08} use the local black-hole mass function and
X-ray constraints on the radiative energy output of AGN to quantify
the accretion and feedback history of AGN within this scenario in a
semi-empirical way.

The HzRGs of our sample have luminous (${\cal L}_{bol}\ge 10^{46}$ erg
s$^{-1}$) AGN as well as very short-lived \citep[a few$\times10^{6-7}$
yrs;] []{blundell99}, powerful radio sources of up to $10^{29}$ W
Hz$^{-1}$ at 500 MHz \citep{miley08}, and they show the signatures of
fast, AGN-driven kpc-scale outflows of significant fractions of their
interstellar medium \citep{nesvadba06a,nesvadba08b}. These are several
of the characteristics expected for the 'flaring' phase between
radiative and kinetic feedback in the 'synthesis' models, and the low
accretion rates of their black holes could be
another. \citet{churazov05} postulate that during this phase, kinetic
feedback and gas heating become {\it more efficient} as accretion
rates drop further, until AGN and host galaxy reach a state where very
low accretion rates can efficiently balance gas cooling
\citep{croton06}. This may be the case in nearby, massive, early-type
galaxies with radio sources \citep{best06,hardcastle07} where star
formation is significantly less efficient than in ordinary star-forming
galaxies \citep{nesvadba10}.

\begin{figure}
\centering
\includegraphics[width=0.8\textwidth]{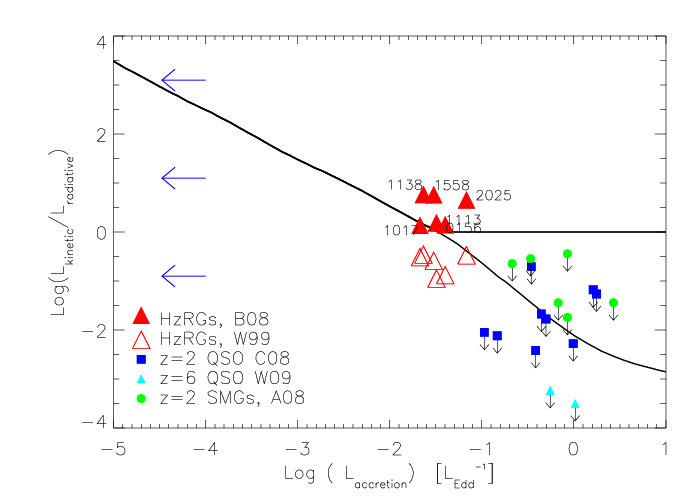}\\
\caption{The ratio of kinetic and radiative luminosity of the AGN as a
function of accretion power, following \citet{merloni08}. Above a
critical accretion power of a few percent of Eddington, the
relationship splits into two branches to accommodate radio-loud and
radio-quiet, rapidly accreting AGN with radiatively efficient
accretion. Red filled and empty triangles show the radio galaxies of
our sample, using different calibrations to estimate the kinetic
energy from the observed radio power (\S\S\ref{ssec:radio},
\ref{sec:discussion.accretion}).  Small blue squares and green circles
show z$\sim$2 quasars and submillimeter galaxies from \citet{coppin08}
and \citet{alexander08}, respectively. Small light blue triangles show
z$\ge$6 quasars taken from \citet{wang10}. Blue arrows mark the
typical accretion regime of massive early-type (and typically
radio-active) elliptical galaxies. Only the HzRGs fall into the
critical accretion regime, where \citep[if the scenario of][and others
is fundamentally correct]{merloni08} we may expect to find evolved
supermassive black holes near the end of their active phase of growth,
as they are transiting into the low-accretion regime spanned by
massive early-type galaxies.}
\label{fig:accretion}
\end{figure}

For a more quantitative comparison we show the ratio of kinetic to
radiative energy output (derived in \S\ref{ssec:radio} and
\S\ref{ssec:accretion}, respectively) in Figure~\ref{fig:accretion} as
a function of accretion luminosity. We adopted this Figure from
\citet[][their Figure~1.]{merloni08}, and use the kinetic luminosities
derived with the relationship of \citet{willott99} and
\citet{birzan08},respectively.
For
accretion luminosities above the critical value, the line splits to
accommodate radio-quiet (lower branch) and radio-loud (upper branch)
AGN. All of our targets are in the critical accretion regime, and have
ratios of kinetic-to-radiative power that are near the values
suggested by \citet{merloni08}.

Figure~\ref{fig:accretion} also shows other populations of powerful
AGN in massive, intensely star-forming galaxies at z$\ge$2, like
submillimeter galaxies with BLR measurements \citep[][]{alexander08},
and quasars at redshift z$\sim$2 and z$\sim$6, respectively
\citep[taken from][]{coppin08, walter04, wang10}. To estimate the
kinetic luminosity of the radio sources in these samples, we used the
approach outlined in \S\ref{ssec:radio} and published radio fluxes,
where we assumed a radio spectral index of $\alpha=-$0.8. For the
high-redshift AGN, this has the character of an upper limit, because
the observed faint radio fluxes may be significantly contaminated, or
even dominated, by star formation. Unlike the HzRGs, these galaxies do
not fall into the critical accretion regime. 

Similarly, the black holes in our HzRGs seem to have lower accretion
luminosities than equally massive black holes studied in large quasar
surveys at z$\sim$2-3. For example, \citet{kollmeier06} find that
typical accretion luminosities in their X-ray and infrared-selected
quasar sample taken from the AGES survey are a factor 10 higher (but
tails of their distribution extend to the same range found for
HzRGs). Their selection suggests $>$70\% completeness for black holes
with masses and accretion luminosities $\ge$0.01 ${\cal L}_{Edd}$,
similar to those in HzRGs, making this comparison robust \citep[but
see, e.g.][ for a discussion of the effects of incompleteness in
statistical studies of the accretion history of
AGN]{kelly10,schulze10}.

\section{Implications for high-redshift galaxy evolution}
\label{sec:implications}
Observational evidence is growing that the host galaxies of HzRGs may
be massive galaxies seen during a late stage of their active formation
phase: At all redshifts, radio galaxies are among the brightest
galaxies in the K-band, suggesting they are at the upper end of the
mass function \citep[][see also \citealt{seymour07}]{debreuck01,
willott03}. At redshifts z$>3$, most HzRGs are luminous submillimeter
emitters with intense star formation, whereas only few galaxies are
detected at lower redshifts \citep[][]{archibald01, reuland04}. If
this is not entirely from selection effects, then it may suggest that
HzRGs as an ensemble are transiting from actively star-forming to
passively evolving during this epoch, perhaps as a consequence of AGN
feedback signaled by the extended outflows of warm ionized gas. It is
interesting that this happens during the epoch where the red sequence
seems to emerge \citep{kodama07}.
 
The results of \S\S\ref{sec:bhbulgemass} and
\ref{sec:discussion.accretion} suggest that the black holes in HzRGs
may also be nearing the end of their period of active growth. Compared
to submillimeter galaxies and quasars at z$\ge$2 \citep{alexander08,
coppin08, walter04, wang10}, they fall remarkably close to the local
relationship between black hole and bulge mass, as can be expected if
the evolution of black holes and bulges is driven by the same
mechanism.

Figure~\ref{fig:accretion}
suggests that the accretion rates of HzRGs may have dropped to a
critical value where mechanical feedback begins to dominate
radiative feedback, unlike the black holes in the above mentioned
submillimeter galaxies and quasars, which all have higher accretion
rates. It is tempting to speculate that this transition may also have
triggered the large-scale outflows of the HzRGs, which may sweep up
and heat the ambient cold gas out to scales of 10s of
kpc. Integral-field spectroscopy of luminous, infrared-selected
quasars with moderate radio sources reveals disturbed gas kinematics,
but no evidence of equally strong outflows extending over scales of
10s of kpc \citep[][Polletta et al., in prep.]{alexander10}.

Obviously, radiatively efficient accretion is maintained in HzRGs,
with ${\cal L}_{bol}=$few$\times 10^{46}$ erg s$^{-1}$, but outflow
rates of a few 100 M$_{\odot}$ yr$^{-1}$ of warm ionized gas
\citep[ignoring other gas phases][]{nesvadba06a,nesvadba08b} suggest
that the gas supply for star formation and AGN activity is being
depleted rapidly. Broad UV absorption lines in MRC2025-218
\citep{humphrey08}, reminiscent of broad absorption lines sometimes
observed in quasars, suggest that outflows are also present near the
black hole, making it plausible that, after the radio-loud phase,
neither the supermassive black holes in HzRGs nor their host
galaxies, will have enough ``fuel'' to maintain much more growth,
unless significant amounts of cold gas are supplied, for example
through a major merger. The environments of HzRGs are rich in ionized,
neutral, and molecular gas \citep[][]{villar03,vanojik97,nesvadba09a},
but the analogy with nearby massive galaxy clusters illustrates that
much of this gas can be heated by the AGN \citep[see, e.g.,][and
references therein]{mcnamara07}.

Moderate, subsequent bulge growth may occur through accretion of
satellite galaxies of HzRGs \citep[][]{nesvadba08b,
hatch09}. Ram-pressure stripping, tidal effects, and on-going star
formation will likely have removed much of the gas in the satellites
by the time they merge, in which case galaxy growth would not
necessarily lead to black-hole growth \citep{kazantzidis05}. This is
consistent with structural parameters of nearby massive early-type and
cD galaxies \citep{weinzirl09, schombert87} and could eliminate the
small offset of HzRGs from the local black-hole bulge mass
relationship (if it is not entirely due to systematic effects), or
else change the position of our galaxies by a few dex within the
large scatter of this relationship.

\section{Summary}
\label{sec:summary}
We presented an analysis of the black-hole properties in 6 powerful
radio galaxies at z$\sim$2 (HzRGs), which show broad, luminous
H$\alpha$ line emission that we identify as emission from the AGN
broad-line region (H$\alpha$BLR). This study is part of an ongoing,
systematic analysis of 50 HzRGs with rest-frame optical integral-field
spectroscopy and with the goal of quantifying the role of AGN feedback
for galaxy evolution at high redshift. Our main results are as
follows.

(1) H$\alpha$BLRs in HzRGs are not rare. 6/31 HzRGs at redshifts where
H$\alpha$ can be observed show H$\alpha$BLRs, corresponding to
$\sim$20\% of our sample.  The broad H$\alpha$ lines have
emission-line luminosities of a few $\times 10^{44}$ erg s$^{-1}$ and
line widths of FWHM$\ge$ 10,000 km s$^{-1}$. Broad H$\beta$ lines in 2
cases and X-ray constraints suggest low extinction (A$_V\sim$ 1 mag).

(2) We estimate black-hole masses and bolometric luminosities from the
H$\alpha$ luminosities and line widths, finding masses of a few $\times
10^{9}$ M$_{\odot}$ and up to $10^{10}$ M$_{\odot}$ which is broadly
consistent with the masses of the most massive black holes in the
nearby universe. Bolometric luminosities are a few percent of the
Eddington luminosity, lower than in other populations of massive,
intensely star-forming galaxies and quasars with black holes of
similar mass.

(3) Black-hole and stellar masses suggest that HzRGs fall on the
black-hole bulge mass relationship for local galaxies, formally with a
small offset to lower bulge masses by at most a factor 4. Much of
this offset may be from systematic effects; alternatively, it could
be explained with an extended phase of accretion of (gas-poor)
satellite galaxies.

(4) Relative to the Eddington luminosities, HzRGs have radiative and
(through the radio source) mechanical energy output rates that are in
the ``critical'' range expected by recent ``unifie''' AGN feedback
scenarios. This could imply that they have reached the end of their
period of active growth, if these models are fundamentally correct. We
argue that the same is probably true for the host galaxies of HzRGs,
suggesting that black hole and host galaxy could be terminating their
phase of active growth at the same time.

\section*{Acknowledgments}
We would like to thank the staff at Paranal for having carried out the
observations and the ESO TAC for the generous allocation of observing
time. We thank the referee for comments that helped substantially
improve the paper. NPHN wishes to thank Elmar K\"ording for
interesting discussions about big and small black holes near and far
during a visit in Southampton. NPHN and MDL wish to thank the CNRS for
the continuous financial support.

\bibliographystyle{aa}
\bibliography{hzrg}

\begin{thebibliography}{111}
\expandafter\ifx\csname natexlab\endcsname\relax\def\natexlab#1{#1}\fi

\bibitem[{{Alexander} {et~al.}(2008){Alexander}, {Brandt}, {Smail}, {Swinbank},
  {Bauer}, {Blain}, {Chapman}, {Coppin}, {Ivison}, \&
  {Men{\'e}ndez-Delmestre}}]{alexander08}
{Alexander}, D.~M., {Brandt}, W.~N., {Smail}, I., {et~al.} 2008, \aj, 135, 1968

\bibitem[{{Alexander} {et~al.}(2010){Alexander}, {Swinbank}, {Smail},
  {McDermid}, \& {Nesvadba}}]{alexander10}
{Alexander}, D.~M., {Swinbank}, A.~M., {Smail}, I., {McDermid}, R., \&
  {Nesvadba}, N.~P.~H. 2010, \mnras, 57

\bibitem[{{Antonucci}(1993)}]{antonucci93}
{Antonucci}, R. 1993, \araa, 31, 473

\bibitem[{{Archibald} {et~al.}(2001){Archibald}, {Dunlop}, {Hughes},
  {Rawlings}, {Eales}, \& {Ivison}}]{archibald01}
{Archibald}, E.~N., {Dunlop}, J.~S., {Hughes}, D.~H., {et~al.} 2001, \mnras,
  323, 417

\bibitem[{{Best} {et~al.}(2006){Best}, {Kaiser}, {Heckman}, \&
  {Kauffmann}}]{best06}
{Best}, P.~N., {Kaiser}, C.~R., {Heckman}, T.~M., \& {Kauffmann}, G. 2006,
  \mnras, 368, L67

\bibitem[{{Bicknell}(1995)}]{bicknell95}
{Bicknell}, G.~V. 1995, \apjs, 101, 29

\bibitem[{{B{\^i}rzan} {et~al.}(2008){B{\^i}rzan}, {McNamara}, {Nulsen},
  {Carilli}, \& {Wise}}]{birzan08}
{B{\^i}rzan}, L., {McNamara}, B.~R., {Nulsen}, P.~E.~J., {Carilli}, C.~L., \&
  {Wise}, M.~W. 2008, \apj, 686, 859

\bibitem[{{Blundell} \& {Rawlings}(1999)}]{blundell99}
{Blundell}, K.~M. \& {Rawlings}, S. 1999, \nat, 399, 330

\bibitem[{{Bonnet} {et~al.}(2004){Bonnet}, {Abuter}, {Baker}, {Bornemann},
  {Brown}, {Castillo}, {Conzelmann}, {Damster}, {Davies}, {Delabre},
  {Donaldson}, {Dumas}, {Eisenhauer}, {Elswijk}, {Fedrigo}, {Finger},
  {Gemperlein}, {Genzel}, {Gilbert}, {Gillet}, {Goldbrunner}, {Horrobin}, {Ter
  Horst}, {Huber}, {Hubin}, {Iserlohe}, {Kaufer}, {Kissler-Patig}, {Kragt},
  {Kroes}, {Lehnert}, {Lieb}, {Liske}, {Lizon}, {Lutz}, {Modigliani}, {Monnet},
  {Nesvadba}, {Patig}, {Pragt}, {Reunanen}, {R{\"o}hrle}, {Rossi}, {Schmutzer},
  {Schoenmaker}, {Schreiber}, {Stroebele}, {Szeifert}, {Tacconi}, {Tecza},
  {Thatte}, {Tordo}, {van der Werf}, \& {Weisz}}]{bonnet04}
{Bonnet}, H., {Abuter}, R., {Baker}, A., {et~al.} 2004, The Messenger, 117, 17

\bibitem[{{Booth} \& {Schaye}(2010)}]{booth09}
{Booth}, C.~M. \& {Schaye}, J. 2010, \mnras, 405, L1

\bibitem[{{Bower} {et~al.}(2006){Bower}, {Benson}, {Malbon}, {Helly}, {Frenk},
  {Baugh}, {Cole}, \& {Lacey}}]{bower06}
{Bower}, R.~G., {Benson}, A.~J., {Malbon}, R., {et~al.} 2006, \mnras, 370, 645

\bibitem[{{Caputi} {et~al.}(2006){Caputi}, {Dole}, {Lagache}, {McLure},
  {Puget}, {Rieke}, {Dunlop}, {Le Floc'h}, {Papovich}, \&
  {P{\'e}rez-Gonz{\'a}lez}}]{caputi06}
{Caputi}, K.~I., {Dole}, H., {Lagache}, G., {et~al.} 2006, \apj, 637, 727

\bibitem[{{Carilli} {et~al.}(2002){Carilli}, {Harris}, {Pentericci},
  {R{\"o}ttgering}, {Miley}, {Kurk}, \& {van Breugel}}]{carilli02}
{Carilli}, C.~L., {Harris}, D.~E., {Pentericci}, L., {et~al.} 2002, \apj, 567,
  781

\bibitem[{{Carilli} {et~al.}(1997){Carilli}, {R\"ottgering}, {van Ojik},
  {Miley}, \& {van Breugel}}]{carilli97}
{Carilli}, C.~L., {R\"ottgering}, H.~J.~A., {van Ojik}, R., {Miley}, G.~K., \&
  {van Breugel}, W.~J.~M. 1997, \apjs, 109, 1

\bibitem[{{Cattaneo} {et~al.}(2007){Cattaneo}, {Blaizot}, {Weinberg}, {Kere{\v
  s}}, {Colombi}, {Dav{\'e}}, {Devriendt}, {Guiderdoni}, \&
  {Katz}}]{cattaneo07}
{Cattaneo}, A., {Blaizot}, J., {Weinberg}, D.~H., {et~al.} 2007, \mnras, 377,
  63

\bibitem[{{Cattaneo} {et~al.}(2009){Cattaneo}, {Faber}, {Binney}, {Dekel},
  {Kormendy}, {Mushotzky}, {Babul}, {Best}, {Br{\"u}ggen}, {Fabian}, {Frenk},
  {Khalatyan}, {Netzer}, {Mahdavi}, {Silk}, {Steinmetz}, \&
  {Wisotzki}}]{cattaneo09}
{Cattaneo}, A., {Faber}, S.~M., {Binney}, J., {et~al.} 2009, \nat, 460, 213

\bibitem[{{Churazov} {et~al.}(2005){Churazov}, {Sazonov}, {Sunyaev}, {Forman},
  {Jones}, \& {B{\"o}hringer}}]{churazov05}
{Churazov}, E., {Sazonov}, S., {Sunyaev}, R., {et~al.} 2005, \mnras, 363, L91

\bibitem[{{Ciotti} \& {Ostriker}(2007)}]{ciotti07}
{Ciotti}, L. \& {Ostriker}, J.~P. 2007, \apj, 665, 1038

\bibitem[{{Ciotti} {et~al.}(2009){Ciotti}, {Ostriker}, \& {Proga}}]{ciotti09}
{Ciotti}, L., {Ostriker}, J.~P., \& {Proga}, D. 2009, \apj, 699, 89

\bibitem[{{Collin} {et~al.}(2006){Collin}, {Kawaguchi}, {Peterson}, \&
  {Vestergaard}}]{collin06}
{Collin}, S., {Kawaguchi}, T., {Peterson}, B.~M., \& {Vestergaard}, M. 2006,
  \aap, 456, 75

\bibitem[{{Coppin} {et~al.}(2008){Coppin}, {Swinbank}, {Neri}, {Cox},
  {Alexander}, {Smail}, {Page}, {Stevens}, {Knudsen}, {Ivison}, {Beelen},
  {Bertoldi}, \& {Omont}}]{coppin08}
{Coppin}, K.~E.~K., {Swinbank}, A.~M., {Neri}, R., {et~al.} 2008, \mnras, 389,
  45

\bibitem[{{Croton} {et~al.}(2006){Croton}, {Springel}, {White}, {De Lucia},
  {Frenk}, {Gao}, {Jenkins}, {Kauffmann}, {Navarro}, \& {Yoshida}}]{croton06}
{Croton}, D.~J., {Springel}, V., {White}, S.~D.~M., {et~al.} 2006, \mnras, 365,
  11

\bibitem[{{De Breuck} {et~al.}(2005){De Breuck}, {Downes}, {Neri}, {van
  Breugel}, {Reuland}, {Omont}, \& {Ivison}}]{debreuck05}
{De Breuck}, C., {Downes}, D., {Neri}, R., {et~al.} 2005, \aap, 430, L1

\bibitem[{{De Breuck} {et~al.}(2003){De Breuck}, {Neri}, \&
  {Omont}}]{debreuck03}
{De Breuck}, C., {Neri}, R., \& {Omont}, A. 2003, New Astronomy Review, 47, 285

\bibitem[{{De Breuck} {et~al.}(2001){De Breuck}, {van Breugel},
  {R{\"o}ttgering}, {Stern}, {Miley}, {de Vries}, {Stanford}, {Kurk}, \&
  {Overzier}}]{debreuck01}
{De Breuck}, C., {van Breugel}, W., {R{\"o}ttgering}, H., {et~al.} 2001, \aj,
  121, 1241

\bibitem[{{Di Matteo} {et~al.}(2005){Di Matteo}, {Springel}, \&
  {Hernquist}}]{dimatteo05}
{Di Matteo}, T., {Springel}, V., \& {Hernquist}, L. 2005, \nat, 433, 604

\bibitem[{{Dong} {et~al.}(2008){Dong}, {Wang}, {Wang}, {Yuan}, {Zhou}, {Dai},
  \& {Zhang}}]{dong08}
{Dong}, X., {Wang}, T., {Wang}, J., {et~al.} 2008, \mnras, 383, 581

\bibitem[{{Eales} \& {Rawlings}(1996)}]{eales96}
{Eales}, S.~A. \& {Rawlings}, S. 1996, \apj, 460, 68

\bibitem[{{Economou} {et~al.}(1995){Economou}, {Lawrence}, {Ward}, \&
  {Blanco}}]{economou95}
{Economou}, F., {Lawrence}, A., {Ward}, M.~J., \& {Blanco}, P.~R. 1995, \mnras,
  272, L5

\bibitem[{{Fanidakis} {et~al.}(2009){Fanidakis}, {Baugh}, {Benson}, {Bower},
  {Cole}, {Done}, \& {Frenk}}]{fanidakis09}
{Fanidakis}, N., {Baugh}, C.~M., {Benson}, A.~J., {et~al.} 2009, ArXiv e-prints

\bibitem[{{Fender} {et~al.}(2004){Fender}, {Belloni}, \& {Gallo}}]{fender04}
{Fender}, R.~P., {Belloni}, T.~M., \& {Gallo}, E. 2004, \mnras, 355, 1105

\bibitem[{{Ferrarese}(2002)}]{ferrarese02}
{Ferrarese}, L. 2002, \apj, 578, 90

\bibitem[{{Gebhardt} {et~al.}(2000){Gebhardt}, {Bender}, {Bower}, {Dressler},
  {Faber}, {Filippenko}, {Green}, {Grillmair}, {Ho}, {Kormendy}, {Lauer},
  {Magorrian}, {Pinkney}, {Richstone}, \& {Tremaine}}]{gebhardt00}
{Gebhardt}, K., {Bender}, R., {Bower}, G., {et~al.} 2000, \apjl, 539, L13

\bibitem[{{Graham} {et~al.}(2001){Graham}, {Erwin}, {Caon}, \&
  {Trujillo}}]{graham01}
{Graham}, A.~W., {Erwin}, P., {Caon}, N., \& {Trujillo}, I. 2001, \apjl, 563,
  L11

\bibitem[{{Greene} \& {Ho}(2005)}]{greene05}
{Greene}, J.~E. \& {Ho}, L.~C. 2005, \apj, 630, 122

\bibitem[{{Hardcastle} {et~al.}(2007){Hardcastle}, {Evans}, \&
  {Croston}}]{hardcastle07}
{Hardcastle}, M.~J., {Evans}, D.~A., \& {Croston}, J.~H. 2007, \mnras, 376,
  1849

\bibitem[{{H{\"a}ring} \& {Rix}(2004)}]{haring04}
{H{\"a}ring}, N. \& {Rix}, H.-W. 2004, \apjl, 604, L89

\bibitem[{{Hatch} {et~al.}(2009){Hatch}, {Overzier}, {Kurk}, {Miley},
  {R{\"o}ttgering}, \& {Zirm}}]{hatch09}
{Hatch}, N.~A., {Overzier}, R.~A., {Kurk}, J.~D., {et~al.} 2009, \mnras, 395,
  114

\bibitem[{{Hopkins} {et~al.}(2006){Hopkins}, {Hernquist}, {Cox}, {Di Matteo},
  {Robertson}, \& {Springel}}]{hopkins06}
{Hopkins}, P.~F., {Hernquist}, L., {Cox}, T.~J., {et~al.} 2006, \apjs, 163, 1

\bibitem[{{Hopkins} {et~al.}(2007){Hopkins}, {Hernquist}, {Cox}, {Robertson},
  \& {Krause}}]{hopkins07}
{Hopkins}, P.~F., {Hernquist}, L., {Cox}, T.~J., {Robertson}, B., \& {Krause},
  E. 2007, \apj, 669, 45

\bibitem[{{Humphrey} {et~al.}(2007){Humphrey}, {Villar-Mart{\'{\i}}n},
  {Fosbury}, {Binette}, {Vernet}, {De Breuck}, \& {di Serego
  Alighieri}}]{humphrey07}
{Humphrey}, A., {Villar-Mart{\'{\i}}n}, M., {Fosbury}, R., {et~al.} 2007,
  \mnras, 375, 705

\bibitem[{{Humphrey} {et~al.}(2008){Humphrey}, {Villar-Mart{\'{\i}}n},
  {Vernet}, {Fosbury}, {di Serego Alighieri}, \& {Binette}}]{humphrey08}
{Humphrey}, A., {Villar-Mart{\'{\i}}n}, M., {Vernet}, J., {et~al.} 2008,
  \mnras, 383, 11

\bibitem[{{Imanishi} \& {Terashima}(2004)}]{imanishi04}
{Imanishi}, M. \& {Terashima}, Y. 2004, \aj, 127, 758

\bibitem[{{Iwamuro} {et~al.}(2002){Iwamuro}, {Motohara}, {Maihara}, {Kimura},
  {Yoshii}, \& {Doi}}]{iwamuro02}
{Iwamuro}, F., {Motohara}, K., {Maihara}, T., {et~al.} 2002, \apj, 565, 63

\bibitem[{{Jarvis} \& {McLure}(2002)}]{jarvis02}
{Jarvis}, M.~J. \& {McLure}, R.~J. 2002, \mnras, 336, L38

\bibitem[{{Kaspi} {et~al.}(2000){Kaspi}, {Smith}, {Netzer}, {Maoz}, {Jannuzi},
  \& {Giveon}}]{kaspi00}
{Kaspi}, S., {Smith}, P.~S., {Netzer}, H., {et~al.} 2000, \apj, 533, 631

\bibitem[{{Kazantzidis} {et~al.}(2005){Kazantzidis}, {Mayer}, {Colpi}, {Madau},
  {Debattista}, {Wadsley}, {Stadel}, {Quinn}, \& {Moore}}]{kazantzidis05}
{Kazantzidis}, S., {Mayer}, L., {Colpi}, M., {et~al.} 2005, \apjl, 623, L67

\bibitem[{{Kelly} {et~al.}(2010){Kelly}, {Vestergaard}, {Fan}, {Hopkins},
  {Hernquist}, \& {Siemiginowska}}]{kelly10}
{Kelly}, B.~C., {Vestergaard}, M., {Fan}, X., {et~al.} 2010, \apj, 719, 1315

\bibitem[{{Kodama} {et~al.}(2007){Kodama}, {Tanaka}, {Kajisawa}, {Kurk},
  {Venemans}, {De Breuck}, {Vernet}, \& {Lidman}}]{kodama07}
{Kodama}, T., {Tanaka}, I., {Kajisawa}, M., {et~al.} 2007, \mnras, 377, 1717

\bibitem[{{Kollmeier} {et~al.}(2006){Kollmeier}, {Onken}, {Kochanek}, {Gould},
  {Weinberg}, {Dietrich}, {Cool}, {Dey}, {Eisenstein}, {Jannuzi}, {Le Floc'h},
  \& {Stern}}]{kollmeier06}
{Kollmeier}, J.~A., {Onken}, C.~A., {Kochanek}, C.~S., {et~al.} 2006, \apj,
  648, 128

\bibitem[{{K{\"o}rding} {et~al.}(2006{\natexlab{a}}){K{\"o}rding}, {Fender}, \&
  {Migliari}}]{koerding06b}
{K{\"o}rding}, E.~G., {Fender}, R.~P., \& {Migliari}, S. 2006{\natexlab{a}},
  \mnras, 369, 1451

\bibitem[{{K{\"o}rding} {et~al.}(2006{\natexlab{b}}){K{\"o}rding}, {Jester}, \&
  {Fender}}]{koerding06a}
{K{\"o}rding}, E.~G., {Jester}, S., \& {Fender}, R. 2006{\natexlab{b}}, \mnras,
  372, 1366

\bibitem[{{K{\"o}rding} {et~al.}(2008){K{\"o}rding}, {Jester}, \&
  {Fender}}]{koerding08}
{K{\"o}rding}, E.~G., {Jester}, S., \& {Fender}, R. 2008, \mnras, 383, 277

\bibitem[{{Kurosawa} \& {Proga}(2009)}]{kurosawa09}
{Kurosawa}, R. \& {Proga}, D. 2009, \mnras, 397, 1791

\bibitem[{{Larkin} {et~al.}(2000){Larkin}, {McLean}, {Graham}, {Becklin},
  {Figer}, {Gilbert}, {Levenson}, {Teplitz}, {Wilcox}, \&
  {Glassman}}]{larkin00}
{Larkin}, J.~E., {McLean}, I.~S., {Graham}, J.~R., {et~al.} 2000, \apjl, 533,
  L61

\bibitem[{{Lauer} {et~al.}(2007){Lauer}, {Tremaine}, {Richstone}, \&
  {Faber}}]{lauer07}
{Lauer}, T.~R., {Tremaine}, S., {Richstone}, D., \& {Faber}, S.~M. 2007, \apj,
  670, 249

\bibitem[{{Maccarone} {et~al.}(2003){Maccarone}, {Gallo}, \&
  {Fender}}]{maccarone03}
{Maccarone}, T.~J., {Gallo}, E., \& {Fender}, R. 2003, \mnras, 345, L19

\bibitem[{{Magorrian} {et~al.}(1998){Magorrian}, {Tremaine}, {Richstone},
  {Bender}, {Bower}, {Dressler}, {Faber}, {Gebhardt}, {Green}, {Grillmair},
  {Kormendy}, \& {Lauer}}]{magorrian98}
{Magorrian}, J., {Tremaine}, S., {Richstone}, D., {et~al.} 1998, \aj, 115, 2285

\bibitem[{{McLure} \& {Jarvis}(2002)}]{mclure02}
{McLure}, R.~J. \& {Jarvis}, M.~J. 2002, \mnras, 337, 109

\bibitem[{{McLure} {et~al.}(2006){McLure}, {Jarvis}, {Targett}, {Dunlop}, \&
  {Best}}]{mclure06}
{McLure}, R.~J., {Jarvis}, M.~J., {Targett}, T.~A., {Dunlop}, J.~S., \& {Best},
  P.~N. 2006, \mnras, 368, 1395

\bibitem[{{McNamara} \& {Nulsen}(2007)}]{mcnamara07}
{McNamara}, B.~R. \& {Nulsen}, P.~E.~J. 2007, \araa, 45, 117

\bibitem[{{Merloni} \& {Heinz}(2008)}]{merloni08}
{Merloni}, A. \& {Heinz}, S. 2008, \mnras, 388, 1011

\bibitem[{{Miley} \& {De Breuck}(2008)}]{miley08}
{Miley}, G. \& {De Breuck}, C. 2008, \aapr, 1

\bibitem[{{Nesvadba} {et~al.}(2010){Nesvadba}, {Boulanger}, {Salome},
  {et~al.}}]{nesvadba10}
{Nesvadba}, N., {Boulanger}, F., {Salome}, P., {et~al.} 2010, \aap, accepted

\bibitem[{{Nesvadba} {et~al.}(2008{\natexlab{a}}){Nesvadba}, {Lehnert},
  {Davies}, {Verma}, \& {Eisenhauer}}]{nesvadba08}
{Nesvadba}, N.~P.~H., {Lehnert}, M.~D., {Davies}, R.~I., {Verma}, A., \&
  {Eisenhauer}, F. 2008{\natexlab{a}}, \aap, 479, 67

\bibitem[{{Nesvadba} {et~al.}(2007){Nesvadba}, {Lehnert}, {De Breuck},
  {Gilbert}, \& {van Breugel}}]{nesvadba07b}
{Nesvadba}, N.~P.~H., {Lehnert}, M.~D., {De Breuck}, C., {Gilbert}, A., \& {van
  Breugel}, W. 2007, \aap, 475, 145

\bibitem[{{Nesvadba} {et~al.}(2008{\natexlab{b}}){Nesvadba}, {Lehnert}, {De
  Breuck}, {Gilbert}, \& {van Breugel}}]{nesvadba08b}
{Nesvadba}, N.~P.~H., {Lehnert}, M.~D., {De Breuck}, C., {Gilbert}, A.~M., \&
  {van Breugel}, W. 2008{\natexlab{b}}, \aap, 491, 407

\bibitem[{{Nesvadba} {et~al.}(2006){Nesvadba}, {Lehnert}, {Eisenhauer},
  {Gilbert}, {Tecza}, \& {Abuter}}]{nesvadba06a}
{Nesvadba}, N.~P.~H., {Lehnert}, M.~D., {Eisenhauer}, F., {et~al.} 2006, \apj,
  650, 693

\bibitem[{{Nesvadba} {et~al.}(2009){Nesvadba}, {Neri}, {De Breuck}, {Lehnert},
  {Downes}, {Walter}, {Omont}, {Boulanger}, \& {Seymour}}]{nesvadba09a}
{Nesvadba}, N.~P.~H., {Neri}, R., {De Breuck}, C., {et~al.} 2009, \mnras, 395,
  L16

\bibitem[{{Nowak}(1995)}]{nowak95}
{Nowak}, M.~A. 1995, \pasp, 107, 1207

\bibitem[{{Onken} {et~al.}(2004){Onken}, {Ferrarese}, {Merritt}, {Peterson},
  {Pogge}, {Vestergaard}, \& {Wandel}}]{onken04}
{Onken}, C.~A., {Ferrarese}, L., {Merritt}, D., {et~al.} 2004, \apj, 615, 645

\bibitem[{{Overzier} {et~al.}(2005){Overzier}, {Harris}, {Carilli},
  {Pentericci}, {R{\"o}ttgering}, \& {Miley}}]{overzier05}
{Overzier}, R.~A., {Harris}, D.~E., {Carilli}, C.~L., {et~al.} 2005, \aap, 433,
  87

\bibitem[{{Papadopoulos} {et~al.}(2000){Papadopoulos}, {R{\"o}ttgering}, {van
  der Werf}, {Guilloteau}, {Omont}, {van Breugel}, \&
  {Tilanus}}]{papadopoulos00}
{Papadopoulos}, P.~P., {R{\"o}ttgering}, H.~J.~A., {van der Werf}, P.~P.,
  {et~al.} 2000, \apj, 528, 626

\bibitem[{{Pei}(1995)}]{pei95}
{Pei}, Y.~C. 1995, \apj, 438, 623

\bibitem[{{Peng} {et~al.}(2006){Peng}, {Impey}, {Rix}, {Kochanek}, {Keeton},
  {Falco}, {Leh{\'a}r}, \& {McLeod}}]{peng06}
{Peng}, C.~Y., {Impey}, C.~D., {Rix}, H., {et~al.} 2006, \apj, 649, 616

\bibitem[{{Pentericci} {et~al.}(2001){Pentericci}, {McCarthy},
  {R{\"o}ttgering}, {Miley}, {van Breugel}, \& {Fosbury}}]{pentericci01}
{Pentericci}, L., {McCarthy}, P.~J., {R{\"o}ttgering}, H.~J.~A., {et~al.} 2001,
  \apjs, 135, 63

\bibitem[{{Pentericci} {et~al.}(2000){Pentericci}, {Van Reeven}, {Carilli},
  {R{\"o}ttgering}, \& {Miley}}]{pentericci00}
{Pentericci}, L., {Van Reeven}, W., {Carilli}, C.~L., {R{\"o}ttgering},
  H.~J.~A., \& {Miley}, G.~K. 2000, \aaps, 145, 121

\bibitem[{{Peterson} {et~al.}(2004){Peterson}, {Ferrarese}, {Gilbert}, {Kaspi},
  {Malkan}, {Maoz}, {Merritt}, {Netzer}, {Onken}, {Pogge}, {Vestergaard}, \&
  {Wandel}}]{peterson04}
{Peterson}, B.~M., {Ferrarese}, L., {Gilbert}, K.~M., {et~al.} 2004, \apj, 613,
  682

\bibitem[{{Proga} \& {Kallman}(2004)}]{proga04}
{Proga}, D. \& {Kallman}, T.~R. 2004, \apj, 616, 688

\bibitem[{{Rees}(1984)}]{rees84}
{Rees}, M.~J. 1984, \araa, 22, 471

\bibitem[{{Reuland} {et~al.}(2004){Reuland}, {R{\"o}ttgering}, {van Breugel},
  \& {De Breuck}}]{reuland04}
{Reuland}, M., {R{\"o}ttgering}, H., {van Breugel}, W., \& {De Breuck}, C.
  2004, \mnras, 353, 377

\bibitem[{{R\"ottgering} {et~al.}(1997){R\"ottgering}, {van Ojik}, {Miley},
  {Chambers}, {van Breugel}, \& {de Koff}}]{rottgering97}
{R\"ottgering}, H.~J.~A., {van Ojik}, R., {Miley}, G.~K., {et~al.} 1997, \aap,
  326, 505

\bibitem[{{Schombert}(1987)}]{schombert87}
{Schombert}, J.~M. 1987, \apjs, 64, 643

\bibitem[{{Schulze} \& {Wisotzki}(2010)}]{schulze10}
{Schulze}, A. \& {Wisotzki}, L. 2010, \aap, 516, A87+

\bibitem[{{Seymour} {et~al.}(2007){Seymour}, {Stern}, {De Breuck}, {Vernet},
  {Rettura}, {Dickinson}, {Dey}, {Eisenhardt}, {Fosbury}, {Lacy}, {McCarthy},
  {Miley}, {Rocca-Volmerange}, {R{\"o}ttgering}, {Stanford}, {Teplitz}, {van
  Breugel}, \& {Zirm}}]{seymour07}
{Seymour}, N., {Stern}, D., {De Breuck}, C., {et~al.} 2007, \apjs, 171, 353

\bibitem[{{Shankar} {et~al.}(2009){Shankar}, {Bernardi}, \&
  {Haiman}}]{shankar08}
{Shankar}, F., {Bernardi}, M., \& {Haiman}, Z. 2009, \apj, 694, 867

\bibitem[{{Shen} {et~al.}(2008){Shen}, {Greene}, {Strauss}, {Richards}, \&
  {Schneider}}]{shen08}
{Shen}, Y., {Greene}, J.~E., {Strauss}, M.~A., {Richards}, G.~T., \&
  {Schneider}, D.~P. 2008, \apj, 680, 169

\bibitem[{{Shields} {et~al.}(2003){Shields}, {Gebhardt}, {Salviander}, {Wills},
  {Xie}, {Brotherton}, {Yuan}, \& {Dietrich}}]{shields03}
{Shields}, G.~A., {Gebhardt}, K., {Salviander}, S., {et~al.} 2003, \apj, 583,
  124

\bibitem[{{Shields} {et~al.}(2006){Shields}, {Menezes}, {Massart}, \& {Vanden
  Bout}}]{shields06}
{Shields}, G.~A., {Menezes}, K.~L., {Massart}, C.~A., \& {Vanden Bout}, P.
  2006, \apj, 641, 683

\bibitem[{{Sijacki} {et~al.}(2007){Sijacki}, {Springel}, {Di Matteo}, \&
  {Hernquist}}]{sijacki07}
{Sijacki}, D., {Springel}, V., {Di Matteo}, T., \& {Hernquist}, L. 2007,
  \mnras, 380, 877

\bibitem[{{Silk} \& {Rees}(1998)}]{silk98}
{Silk}, J. \& {Rees}, M.~J. 1998, \aap, 331, L1

\bibitem[{{Springel} {et~al.}(2005){Springel}, {Di Matteo}, \&
  {Hernquist}}]{springel05}
{Springel}, V., {Di Matteo}, T., \& {Hernquist}, L. 2005, \mnras, 361, 776

\bibitem[{{Tecza} {et~al.}(2000){Tecza}, {Thatte}, {Eisenhauer}, {Mengel},
  {Roehrle}, \& {Bickert}}]{tecza00}
{Tecza}, M., {Thatte}, N.~A., {Eisenhauer}, F., {et~al.} 2000, in Society of
  Photo-Optical Instrumentation Engineers (SPIE) Conference Series, Vol. 4008,
  Society of Photo-Optical Instrumentation Engineers (SPIE) Conference Series,
  ed. {M.~Iye \& A.~F.~Moorwood}, 1344--1350

\bibitem[{{Tremaine} {et~al.}(2002){Tremaine}, {Gebhardt}, {Bender}, {Bower},
  {Dressler}, {Faber}, {Filippenko}, {Green}, {Grillmair}, {Ho}, {Kormendy},
  {Lauer}, {Magorrian}, {Pinkney}, \& {Richstone}}]{tremaine02}
{Tremaine}, S., {Gebhardt}, K., {Bender}, R., {et~al.} 2002, \apj, 574, 740

\bibitem[{{Treu} {et~al.}(2007){Treu}, {Woo}, {Malkan}, \&
  {Blandford}}]{treu07}
{Treu}, T., {Woo}, J.-H., {Malkan}, M.~A., \& {Blandford}, R.~D. 2007, \apj,
  667, 117

\bibitem[{{van Breugel} {et~al.}(1998){van Breugel}, {Stanford}, {Spinrad},
  {Stern}, \& {Graham}}]{vanbreugel98}
{van Breugel}, W.~J.~M., {Stanford}, S.~A., {Spinrad}, H., {Stern}, D., \&
  {Graham}, J.~R. 1998, \apj, 502, 614

\bibitem[{{van Ojik} {et~al.}(1997){van Ojik}, {R\"ottgering}, {Miley}, \&
  {Hunstead}}]{vanojik97}
{van Ojik}, R., {R\"ottgering}, H.~J.~A., {Miley}, G.~K., \& {Hunstead}, R.~W.
  1997, \aap, 317, 358

\bibitem[{{Vestergaard}(2002)}]{vestergaard02}
{Vestergaard}, M. 2002, \apj, 571, 733

\bibitem[{{Vestergaard} \& {Peterson}(2006)}]{peterson06}
{Vestergaard}, M. \& {Peterson}, B.~M. 2006, \apj, 641, 689

\bibitem[{{Villar-Mart{\'{\i}}n} {et~al.}(2003){Villar-Mart{\'{\i}}n},
  {Vernet}, {di Serego Alighieri}, {Fosbury}, {Humphrey}, \&
  {Pentericci}}]{villar03}
{Villar-Mart{\'{\i}}n}, M., {Vernet}, J., {di Serego Alighieri}, S., {et~al.}
  2003, \mnras, 346, 273

\bibitem[{{Walter} {et~al.}(2004){Walter}, {Carilli}, {Bertoldi}, {Menten},
  {Cox}, {Lo}, {Fan}, \& {Strauss}}]{walter04}
{Walter}, F., {Carilli}, C., {Bertoldi}, F., {et~al.} 2004, \apjl, 615, L17

\bibitem[{{Wang} {et~al.}(2010){Wang}, {Carilli}, {Neri}, {Riechers}, {Wagg},
  {Walter}, {Bertoldi}, {Menten}, {Omont}, {Cox}, \& {Fan}}]{wang10}
{Wang}, R., {Carilli}, C.~L., {Neri}, R., {et~al.} 2010, ArXiv e-prints

\bibitem[{{Ward} {et~al.}(1988){Ward}, {Done}, {Fabian}, {Tennant}, \&
  {Shafer}}]{ward88}
{Ward}, M.~J., {Done}, C., {Fabian}, A.~C., {Tennant}, A.~F., \& {Shafer},
  R.~A. 1988, \apj, 324, 767

\bibitem[{{Weinzirl} {et~al.}(2009){Weinzirl}, {Jogee}, {Khochfar}, {Burkert},
  \& {Kormendy}}]{weinzirl09}
{Weinzirl}, T., {Jogee}, S., {Khochfar}, S., {Burkert}, A., \& {Kormendy}, J.
  2009, \apj, 696, 411

\bibitem[{{Willott} {et~al.}(1999){Willott}, {Rawlings}, {Blundell}, \&
  {Lacy}}]{willott99}
{Willott}, C.~J., {Rawlings}, S., {Blundell}, K.~M., \& {Lacy}, M. 1999,
  \mnras, 309, 1017

\bibitem[{{Willott} {et~al.}(2000){Willott}, {Rawlings}, \&
  {Jarvis}}]{willott00}
{Willott}, C.~J., {Rawlings}, S., \& {Jarvis}, M.~J. 2000, \mnras, 313, 237

\bibitem[{{Willott} {et~al.}(2003){Willott}, {Rawlings}, {Jarvis}, \&
  {Blundell}}]{willott03}
{Willott}, C.~J., {Rawlings}, S., {Jarvis}, M.~J., \& {Blundell}, K.~M. 2003,
  \mnras, 339, 173

\bibitem[{{Woo} {et~al.}(2006){Woo}, {Treu}, {Malkan}, \& {Blandford}}]{woo06}
{Woo}, J., {Treu}, T., {Malkan}, M.~A., \& {Blandford}, R.~D. 2006, \apj, 645,
  900

\bibitem[{{Woo} {et~al.}(2008){Woo}, {Treu}, {Malkan}, \& {Blandford}}]{woo08}
{Woo}, J., {Treu}, T., {Malkan}, M.~A., \& {Blandford}, R.~D. 2008, \apj, 681,
  925

\bibitem[{{Wyithe}(2006)}]{wyithe06}
{Wyithe}, J.~S.~B. 2006, \mnras, 365, 1082

\bibitem[{{Younger} {et~al.}(2008){Younger}, {Hopkins}, {Cox}, \&
  {Hernquist}}]{younger08}
{Younger}, J.~D., {Hopkins}, P.~F., {Cox}, T.~J., \& {Hernquist}, L. 2008,
  \apj, 686, 815

\end{thebibliography}

\clearpage
\onecolumn

\clearpage
\onecolumn

\clearpage
\onecolumn

\clearpage
\onecolumn

\clearpage
\onecolumn

\clearpage
\onecolumn
\begin{table}
\begin{minipage}[t]{\columnwidth}
\caption{Observed wavelengths, redshifts, full-width a half maximum, fluxes of the H$\alpha$ and H$\beta$ broad lines, and V-band extinctions.}
\label{tab:observations}
\centering
\renewcommand{\footnoterule}{}  
\begin{tabular}{lcccccc}
\hline
\hline
Source & $\lambda_{\rm obs}$ & Redshift & FWHM(H$\alpha$)\footnote{corrected for instrumental resolution} &  F(H$\alpha$)\footnote{for the broad line} & F(H$\beta$)\footnote{for the broad line}. & A$_V$\footnote{V-band extinction of the broad line region.} \\
       & $[\AA]$            &          & $[$km s$^{-1}]$ & $[$erg s$^{-1}]$ & $[$erg s$^{-1}]$ & $[$mag$]$ \\
\hline
MRC0156-252 & 19764.1       & 2.01143  & 12436  & $1.75\times10^{-14}$   &  &  \\ 
MRC1017-220 & 18202.9       & 1.77356  & 12006  & $5.3\times10^{-15}$    &  &  \\ 
TXS1113-178 & 21253.7       & 2.23851  & 11063  & $1.4\times 10^{-14}$   &  &  \\ 
MRC1138-262 & 20817.9       & 2.17201  & 14900  & $1.9\times10^{-14}$    &  &  \\
MRC1558-003 & 23002.5       & 2.50488  & 12425  & $8.3\times 10^{-15}$   & $2.1\times 10^{-15}$ & 0.9  \\
MRC2025-218 & 23829.9       & 2.63094  & 8023.6 & $5.5\times10^{-15}$    & $1.3\times 10^{-15}$ & 1.0  \\
\hline
\hline
\end{tabular}
\end{minipage}
\end{table}

\begin{table}
\caption{Broad H$\alpha$ luminosities, black holes masses, Eddington and bolometric luminosities, jet kinetic luminosities, and stellar masses.} 
\label{tab:results}
\centering
\renewcommand{\footnoterule}{}  
\begin{minipage}[t]{\columnwidth}
\begin{tabular}{lcccccccc}
\hline
\hline
Source & ${\cal L}$(H$\alpha$)\footnote{not corrected for extinction} & M$_{\rm BH}$\footnote{masses corrected for a putative orientation bias would be factors 2 lower}&  ${\cal L}_{\rm Edd}$\footnote{Eddington luminosities corrected for a putative orientation bias would be factors 2 lower.} & ${\cal L}_{\rm bol}$\footnote{not corrected for extinction} &  ${\cal L}_{kin,B08}$\footnote{estimated with the relationship of \citet{birzan08}} & ${\cal L}_{kin,W99}$\footnote{estimated with the relationship of \citet{willott99}} & M$_{\rm stellar}$\footnote{from \citet{seymour07}} \\
     &        $[$erg s$^{-1}]$           &   $[$M$_{\odot}]$      & $[$erg s$^{-1}]$          & $[$erg s$^{-1}]$  & $[$erg s$^{-1}]$ & $[$erg s$^{-1}]$  & $[$M$_{\odot}]$ \\
\hline
MRC0156-252 & $5.0\times 10^{44}$ & $1\times 10^{10}$ & $1\times 10^{48}$ & $5\times 10^{46}$ & $5\times 10^{45}$ & 7$\times10^{46}$ & $<$1.6$\times10^{12}$ \\ 
MRC1017-220 & $1.2\times 10^{44}$ & $5\times 10^{9 }$ & $6\times 10^{47}$ & $1\times 10^{46}$ & $4\times 10^{45}$ & 2$\times10^{46}$ & $<$5$\times10^{11}$\\ 
TXS1113-178 & $5.3\times 10^{44}$ & $9\times 10^{9 }$ & $1\times 10^{48}$ & $5\times 10^{46}$ & $6\times 10^{45}$ & 7$\times10^{46}$ & N/A \\ 
MRC1138-262 & $6.6\times 10^{44}$ & $2\times 10^{10}$ & $3\times 10^{48}$ & $6\times 10^{46}$ & $2\times 10^{46}$ & 3$\times10^{47}$ & $<$1.2$\times10^{12}$ \\
MRC1558-003 & $4.2\times 10^{44}$ & $1\times 10^{10}$ & $1\times 10^{48}$ & $4\times 10^{46}$ & $9\times 10^{45}$ & 2$\times10^{47}$ & 5$\times10^{11}$ \\
MRC2025-218 & $3.1\times 10^{44}$ & $3\times 10^{9 }$ & $4\times 10^{47}$ & $3\times 10^{46}$ & $1\times 10^{46}$ & 1$\times10^{47}$ & 4$\times10^{11}$ \\
\hline
\hline
\end{tabular}
\end{minipage}
\end{table}

\end{document}